\documentclass{iopart}

\def\be{\begin{equation}}
\def\ee{\end{equation}}
\def\bea{\begin{eqnarray}}
\def\eea{\end{eqnarray}}

\usepackage{iopams}

\begin{document}
\title[Holographic formula]
{Holographic formula for the determinant of the scattering operator in thermal AdS}

\author{D E D\'{\i}az} 

\address{Universidad Andr\'es Bello,
Departamento de Ciencias F\'{\i}sicas,
Av. Republica 252, Santiago, Chile}
\ead{danilodiaz@unab.cl}

\begin{abstract} A `holographic formula' expressing the functional determinant of the scattering operator in an asymptotically locally anti-de Sitter(ALAdS) space has been proposed in terms of a relative functional determinant of the scalar Laplacian in the bulk. It stems from considerations in AdS/CFT correspondence of a quantum correction to the partition function in the bulk and the corresponding sub-leading correction at large N on the boundary. In this paper we probe this prediction for a class of quotients of hyperbolic space by a discrete subgroup of isometries. We restrict to the simplest situation of an abelian group where the quotient geometry describes thermal AdS and also a non-spinning BTZ instanton. The bulk computation is explicitly done using the method of images and the answer can be encoded in a (Patterson-)Selberg zeta function.
\end{abstract}
\submitto{\JPA}

\section{Introduction}

Maldacena's conjecture~\cite{Malda}, whose current status might be categorized as true but not proven~\cite{HP06}, during its first ten years has laid many ``golden eggs'' and somehow parallels longstanding conjectures in mathematics such as the Riemann hypothesis, which soon will turn hundred-fifty.
There is a vast literature of applications of the correspondence, far beyond the original canonical examples and many of them in the form of gauge/gravity dualities. They range from (AdS-)black holes dual to hot strongly-coupled quark-gluons plasmas (produced by heavy-ion collisions and experimentally accessible at RHIC or LHC) to dual AdS descriptions of cosmological singularities.

Within this broad spectrum, there is also room for interplay with mathematics. For example, one finds that AdS/CFT correspondence has motivated the study of the connectedness of the conformal boundary~\cite{WY99} or the Z-minimization procedure for Sasaki-Einstein manifolds~\cite{MSY06}.
The duality is also very much linked to conformal and asymptotically hyperbolic geometry~\cite{MGC07}. It has revived the successful tradition, initiated by Fefferman and Graham~\cite{FG85}, of addressing the geometry of a conformal manifold as that of the conformal infinity of an asymptotically Einstein metric (although, with hindsight, one can consider the ambient metric of Fefferman and Graham as a curved version of Dirac's conformal space~\cite{Dir36}). Among several AdS/CFT-inspired computations in this area, as relevant to the purposes of this note we should mention the volume renormalization~\cite{Gra99} which has unveiled curvature invariants, alternative definitions of GJMS operators and Q-curvature~\cite{GZ01,FG02}, and a holographic formula expressing the Q-curvature in terms of the coefficients in the asymptotic expansion of the volume~\cite{GJ07}.

Most of these mathematical developments are motivated by AdS/CFT correspondence as a broader principle, as a holographic duality, and by heuristic  results which are sort of universal, in the sense that they do not rely on the specific details (e.g. SUSY, Kaluza-Klein states) of the underlying theory. This is certainly the case for the subject of the present paper, namely a holographic formula relating a relative functional determinant of the Laplacian in a conformally compact asymptotically Einstein space $X^{n+1}$ and the functional determinant of the associated scattering operator at the conformal boundary $M_n$:

\be
\label{hol-for}
\frac{\det_-(\Delta_X-\lambda(n-\lambda))}{\det_+(\Delta_X-\lambda(n-\lambda))}
=\det\,S_M(\lambda)~.
\ee
The (+)-determinant is computed via the Green's function (resolvent) with spectral parameter $\lambda_+$ and the (-)-determinant is obtained by analytic continuation to $\lambda_-$, where $\lambda_+>\lambda_-$ are the roots of the mass/scaling-dimension relation $\lambda(\lambda-n)=m^2$.
As a hint for such a relation, consider the two-point function of a CFT operator $O$ with scaling dimension $\lambda$ and the Green's function of its dual scalar field $\phi$ of mass $m$ and take the mass to be much bigger than the inverse AdS radius (which we set conventionally to 1) so that we have a semiclassical regime(see e.g.~\cite{DiV99}). The Green's function is then dominated by the saddle point action corresponding to the length $\sigma$ (\ref{xi}) of the geodesic connecting the two points
$
G(x,x')\sim e^{-m\sigma}~.
$
In Poincare coordinates, as the two points approach the boundary (both at $z=z'=\epsilon\sim0$), the Green's function behaves as
$
G(x,x')\sim (\parallel\mathbf{x}-\mathbf{x}'\parallel^2/\epsilon^2)^{-m}~.
$
A comparison with the CFT two-point function $\langle
O(\mathbf{x})O(\mathbf{x}')\rangle\sim (\parallel\mathbf{x}-\mathbf{x}'\parallel^2/b^2)^{-\lambda}$, with $b$ an UV cutoff, leads to the identification $m=\lambda$ valid for large $m$ (the precise relation being $m^2=\lambda(\lambda-n)$) and to the IR-UV connection ($\epsilon\sim b$) which maps IR-divergencies in the bulk to UV-divergencies on the boundary, or equivalently, relates the radial coordinate in AdS to the energy scale of the CFT.
Now one can pose a more ambitious question concerning the functional determinant of the two-point function (or scattering operator). The naive comparison with the bulk determinant of the Laplacian with a mass term is not satisfactory because there are too many divergencies in the bulk. The difficulties in regularizing the functional determinant on the boundary are associated to the UV-divergencies, whereas the functional determinant in the bulk has the short distance UV-divergency in addition to the IR-divergence due to the infinite volume of AdS.

The precise holographic relation~(\ref{hol-for}) was reached by a rather circuitous route, involving the one-loop contributions to the partition function in the bulk by a scalar field with mass in the Breitenlohner-Freedman window~\footnote{We consider the continuation to Riemannian or Euclidean signature, mainly because of the CFT side and, in addition, the Penrose compactification of the bulk metric is simpler.} and the corresponding double-trace deformation of the boundary theory~\cite{GM03}. Initially it arose as a prediction for the anomaly from an odd-dimensional hyperbolic bulk, then a confirmation from the boundary theory on the sphere was achieved~\cite{GK03}  as well as a kinematical understanding~\cite{HR06}. Finally, in~\cite{DD07} full agreement was found not only for the anomalous part but also for the renormalized functional determinants in any dimension. In addition, the variation under conformal transformations for $n$ even was explored and revealed a generalized Polyakov formula and the universal coefficient of the type A anomaly for GJMS operators~\cite{DD08}.

In the present work, we explore the interesting case when the bulk $X$ is a quotient of hyperbolic space by a discrete subgroup of isometries. This allows for the inclusion of temperature effects by considering thermal AdS~\cite{HP83}, and as a bonus we also have a non-spinning(Euclidean)BTZ geometry~\cite{BTZ92,BHTZ92}, exploiting their representation as hyperbolic space with identifications. The quotient geometry turns out to be simple enough to allow for analytical calculations and it seems worthwhile to spell out in detail both bulk and boundary computations.
Additionally, one can find rudiments of holography in the mathematical results involving scattering and Selberg zeta function defined for such quotients of hyperbolic space~\cite{Pat89,PP01}. It is our aim to encode all this information, resonances of the scattering operator and zeros of the Selberg zeta function, in the equality involving the functional determinants.

\section{Quotient of hyperbolic space}

Let us start by considering in Poincare's half-space model for hyperbolic space
the identification
\be(z,\mathbf{x})\sim e^l(z,\mathbf{x})~,\ee
which corresponds to an isometry of the metric
\be
g_{X}=\frac{dz^2+d\mathbf{x}^2}{z^2}.\ee

This dilation generates a discrete abelian group $\Gamma$ and one can take as fundamental domain for its action the region $1\leq z \leq e^l$.
The resulting quotient space $X=H^{n+1}/\Gamma$, with topology $S^1(l)\times R^n$, can be written in global coordinates as
\be
g_{X}=(1+r^2)du^2+\frac{dr^2}{1+r^2}+r^2d\Omega^2_{n-1}
\ee
together with the identification $u\sim u+l$, so that the conformal infinity is $S^1(l)\times S^{n-1}$ equipped with the standard metric $du^2+d\Omega^2_{n-1}$.

\subsection{Thermal AdS and non-spinning BTZ instanton}

The physical motivation for considering this simple quotient comes from the fact that AdS as well as
a higher-dimensional non-spinning BTZ black hole, both have Euclidean sections corresponding to the above metric
~\cite{Ban97,BGM98,Cai02,FMRS04}. The choice of $u\sim u+l$ as Euclidean time (inverse temperature $\beta_{TAdS}=l$), after Wick rotation and `unwrapping' to avoid closed timelike curves, results in the AdS metric
\be
g_{AdS}=-(1+r^2)dt^2+\frac{dr^2}{1+r^2}+r^2d\Omega^2_{n-1}.
\ee

The case of the BTZ black hole in turn, corresponds to the choice of an angle on the sphere as Euclidean time. The Wick-rotated metric of the sphere $S^{n-1}$ is taken as
\be
d\Omega^2_{n-1}\rightarrow -d\tau^2+\mbox{ch}^2\tau (d\widetilde{\theta}^2+\mbox{sin}^2
\widetilde{\theta}\,d\Omega^2_{n-3}).
\ee
Notice that the metric is no longer static in dimension higher that three.
The spherical coordinates of ~\cite{BGM98} can be obtained via $\rho^2=1+r^2$, $\mbox{cos}\theta=\mbox{ch}\tau\mbox{sin}\widetilde{\theta}$ and $\mbox{th}t=\mbox{th}\tau/\mbox{cos}\widetilde{\theta}$,
\be
\fl g_{BTZ}=(\rho^2-1)\left[-\mbox{sin}^2 \theta dt^2+d\theta^2+\mbox{cos}^2
\theta\,d\Omega^2_{n-3}\right] + \frac{d\rho^2}{\rho^2-1}+\rho^2du^2.
\ee
A further transformation $\mbox{cos}\theta=\mbox{sh}\chi/\sqrt{\rho^2-1}$ and $\rho=\widetilde{r}\mbox{ch}\chi$, makes explicit a three-dimensional BTZ factor~\cite{FMRS04} where the dependence on $r_+$ has been absorbed in the angular part $u\sim u+2\pi r_+$
\be
\fl g_{BTZ}=\mbox{ch}^2\chi\left[-(\widetilde{r}^2-1)dt^2+
\frac{d\widetilde{r}^2}{\widetilde{r}^2-1}+\widetilde{r}^2 du^2\right] + d\chi^2+
\mbox{sh}^2\chi d\Omega^2_{n-3}.
\ee

In all, the quotient metric describes both thermal AdS with $\beta_{AdS}=l$ and a higher-dimensional version of the non-spinning BTZ instanton with inverse temperature $\beta_{BTZ}=2\pi/r_+=4\pi^2/l$.

\subsection{Dimensionally regularized volume}
Even if the conformal boundary can be made compact, the volume of the fundamental region has infinite hyperbolic volume and one has to deal with its renormalization. To compute (renormalize) the volume of the quotient space $X$, it is convenient to first cast the metric in Fefferman-Graham form~\cite{CQY07} with $r=\frac{1-s^2/4}{s}=\mbox{sh ln}\,\frac{2}{s}$,
\be
g_{X}=s^{-2}\left\{ds^2+(1+s^2/4)^2du^2+(1-s^2/4)^2d\Omega^2_{n-1}\right\}~.
\ee
The computation in dimensional regularization gives an exactly vanishing answer

\be
\label{vol}
\fl
\int d{\it
vol}_X\,=\,{\it vol}_{S^{n-1}}\cdot
\int_0^2ds\,s^{-n-1}\;(1-s^2/4)^{n-1}\;(1+s^2/4)=0.\ee
This is a remarkable result that can probably be traced back to the vanishing Euler characteristic of the circle $\chi(S^1(l))=0$.

\section{Rudiments of holography}

In such a quotient space, one can find rudiments of holography in certain mathematical results involving Selberg zeta function and the scattering operator~\cite{PP01}. There is a correspondence between the zeros (divisor) of Selberg zeta function and the scattering resonances (poles of the meromorphic continuation of the scattering matrix). The zeta function is the bulk-quantity, related to the length of closed geodesics, whereas the scattering operator is essentially a boundary quantity, a two point-correlator.

\subsection{Scattering}
Let us examine the scattering, a straightforward adaptation of the three-dimensional computation by Perry and Williams~\cite{PW03}, starting with the Laplace operator and the metric in ``scattering form''
\be
g_{X}=dt^2+ \mbox{ch}^2t\,du^2+ \mbox{sh}^2t\,d\Omega^2_{n-1}~.
\ee
The positive Laplacian is given by
\be
\Delta_X=-\frac{1}{\mbox{ch}t\,\mbox{sh}^{n-1}t}\,\partial_t\left(\mbox{ch}t\,\mbox{sh}^{n-1}t\,\partial_t\right)
-\frac{1}{\mbox{ch}^2t}\partial_u^2-\frac{1}{\mbox{sh}^2t}\Delta_{\Omega},
\ee
and a Hilbert space decomposition, using spherical harmonics for the angular part and $\exp(2\pi iNu/l) ( N\in \mathbb{Z}$) for the circle, results in the effective one-dimensional operator

\be\fl
\widehat{H}_{L,N}=-\frac{1}{\mbox{ch}t\,\mbox{sh}^{n-1}t}\,d_t\left(\mbox{ch}t\,\mbox{sh}^{n-1}t\,d_t\right)
+\kappa^2_N\,\mbox{sech}^2t+\gamma^2_L\,\mbox{csch}^2t~,
\ee
where $\gamma^2_L=L(L+n-2)$ and $\kappa^2_N=(2\pi N/l)^2$ are the corresponding eigenvalues.
The effective one-dimensional operator, after a unitary transformation to act on functions rescaled by $U^{-1}=\mbox{ch}^{1/2}t\,\mbox{sh}^{(n-1)/2}t$, that is $H_{L,N}=U^{-1}\widehat{H}_{L,N}U$, can be brought into the familiar form of the one-dimensional stationary Schr\"odinger equation with a (generalized) P\"oschl-Teller potential

\be
H_{L,N}=-d^2_t+\alpha(\alpha+1)\,\mbox{csch}^2t-\beta(\beta+1)\,\mbox{sech}^2t
\ee
with $\alpha=-3/2+n/2+L$ and $\beta=-1/2+2\pi i \mid N\mid/l$.

The scattering problem or, equivalently, the determination of the thermal correlator at the boundary reduces then to a familiar quantum-mechanical  problem. The scattering matrix (here $\lambda=n/2+\nu$), up to unimportant factors associated to trivial poles, is known to be given by

\be\fl
S_{L,N}(\lambda)=\frac{\Gamma(L/2+n/4+\nu/2+\pi iN/l)\;\Gamma(L/2+n/4+\nu/2-\pi iN/l)}{\Gamma(L/2+n/4-\nu/2+\pi i N/l)\;\Gamma(L/2+n/4-\nu/2-\pi iN/l)}~.
\ee
Notice the lattice of poles (resonances) at $\lambda=-2k-L+2\pi i N/l$, with $k,L=0,1,2,...$ and $N\in \mathbb{Z}$.

\subsection{Selberg zeta function}
Associated to the quotient space $X$, there is a Selberg zeta function. For the BTZ geometry, for example, it is an ubiquitous quantity~\cite{PW03}:
the effective action of the BTZ instanton, the black hole free energy, functional determinants
appearing in quantum corrections to entropy, etc. are all expressible in terms of $Z_{\Gamma}$.

We first recall ($n=2$) the Selberg zeta function associated to the non-spinning BTZ black hole~\cite{PW03}
\be
\mbox{ln\,Z}_{\Gamma}(\lambda)=-\sum^{\infty}_{N=1}\,\frac{1}{N}\,\frac{e^{-\lambda lN}}{(1-e^{-lN})^2}~.
\ee
Following Patterson's construction~\cite{Pat89} of the Selberg zeta function, Perry and Williams define
\be
\mbox{Z}_{\Gamma}(\lambda)
=\prod^{\infty}_{k_1=0}\,\prod^{\infty}_{k_2=0}(1-e^{-(k_1+k_2+\lambda)l})~.
\ee
We can straightforwardly generalize from three to $(n+1)$-dimensional Poincare patch:
\be
\mbox{Z}_{\Gamma}(\lambda)
=\prod^{\infty}_{k_1,k_2,...,k_n=0}\,
(1-e^{-(k_1+k_2+...+k_n+\lambda)l})~.
\ee
The location of the zeros at $\lambda=-k_1-...-k_n+2\pi i N/l$ can be mapped to the lattice of scattering resonances of before.

Taking the logarithm and Taylor-expanding, the resulting geometric series can be summed up
\be\fl
-\sum^{\infty}_{N=1}\,\frac{1}{N}\,\sum^{\infty}_{k_1,k_2,...,k_n=0}\,
e^{-\lambda lN -k_1lN-k_2lN-...-k_nlN}
=-\sum^{\infty}_{N=1}\,\frac{1}{N}\,\frac{e^{-\lambda lN}}{(1-e^{-lN})^n}~,
\ee
so that we have a Selberg zeta function associated to the dilations in the Poincare patch
\be
\mbox{ln\,Z}_{\Gamma}(\lambda)=-\sum^{\infty}_{N=1}\,\frac{1}{N}\,\frac{e^{-\lambda lN}}{(1-e^{-lN})^n}~.
\ee

\section{Boundary: functional determinant of the scattering operator}
Having at hand a Selberg zeta function attached to the identifications by dilations in the $n+1$-dimensional Poincare patch, we now proceed to the direct evaluation of the functional determinant of the scattering operator, this is just the RHS of the holographic formula~(\ref{hol-for}):
\be
\mbox{ln\,det}\,S(\lambda)=\sum^{\infty}_{N=-\infty}\,\sum^{\infty}_{L=0}\,\mbox{deg}(n-1,L)\;
\mbox{ln}\,S_{L,N}~.
\ee
The computation is very similar to the one for the sphere in~\cite{DD07}. Dimensional regularization washes away terms in the sum independent of $L$ and the new feature is the additional periodic direction which brings in an additional sum over $N$ (temperature!). It is convenient to take the derivative with respect to $\lambda$, i.e. to compute in fact $\mbox{tr}(S^{-1}\partial_{\lambda}S(\lambda))$ and using an integral representation for the digamma function $\psi$, as in~\cite{DD07}, one gets
\bea\fl
\mbox{tr}(S^{-1}\partial_{\lambda}S(\lambda))=-\sum^{\infty}_{N=-\infty}\,\sum^{\infty}_{L=0}\,\mbox{deg}(n-1,L)\;\times\\
\times\;\int^{\infty}_0 dt \frac{e^{-tL/2-tn/4-it\pi N/l}}{1-e^{-t}}(e^{t\nu/2}+e^{-t\nu/2})
\eea
The sum over degeneracies is easy to compute
\be
\sum^{\infty}_{L=0}\mbox{deg}(n-1,L)\,e^{-tL/2}=\frac{1+e^{-t/2}}{(1-e^{-t/2})^{n-1}}~.
\ee
In the sum corresponding to the l-periodic $u$ direction we recognize the Fourier transform of ``Dirac's comb'', that is, Poisson summation formula \be
\sum^{\infty}_{N=-\infty}e^{-i\pi tN/l}=2l\sum^{\infty}_{N=-\infty}\delta(t-2lN).
\ee
 Now, the remaining integral is direct to compute, and for $n$ negative enough 
 the only contributions come from positive $N$:
\be
\mbox{tr}(S^{-1}\partial_{\lambda}S(\lambda))=-2l\sum^{\infty}_{N=1}e^{-nlN/2}\,\frac{e^{\nu lN}+e^{-\nu lN}}{(1-e^{-lN})^n}
\ee

Integrating back in $\nu$, or equivalently in $\lambda$, one finally gets for the functional determinant
\be
\mbox{ln\,det}\,S(\lambda)=2\sum^{\infty}_{N=1}\,\frac{1}{N}\,\frac{e^{-\lambda lN}-e^{-(n-\lambda)\,lN}}{(1-e^{-lN})^n}~,
\ee
a result that in terms of the Selberg zeta function can be compactly (and remarkably!) written as
\be
\label{scat-n}
\mbox{det}\,S(\lambda)=\left[\mbox{Z}_{\Gamma}(n-\lambda)/\mbox{Z}_{\Gamma}(\lambda)\right]^2~.
\ee

\section{Bulk: method of images}

Now we want to make the holographic computation, to compute the relative determinant via the AdS/CFT recipe, i.e. the LHS of the holographic formula~(\ref{hol-for}).
The method of images amounts to consider the sum over image locations, lattice of identified points, to produce the usual short-distance singularity as well as the appropriate periodicity of the bulk geometry.
As explained in~\cite{Cam91}, the $+$branch is the usual one that can be obtained e.g. via heat kernel or Green's function techniques, while the $-$brach is obtained by analytic continuation from $\lambda$ to $n-\lambda$. To parallel the boundary computation, we take the derivative with respect to $\lambda$ and compute in terms of the (truncated) heat kernel representation for the Green's function subtracting the direct contribution given by the Green's function for the original hyperbolic space. The trace here means taking the coincidence-point limit and integrating over the fundamental domain, so that
\be\fl
(2\lambda-n)\mbox{tr}{(G^+_X-G^+_H)}=(2\lambda-n)\int^{\infty}_0 ds\,\mbox{tr}\,K'_X(\sigma,s)\,e^{s\lambda(n-\lambda)}~.
\ee

Let us begin three dimensions ($n=2$). Here, Mann and Solodukhin~\cite{MS96} have already computed the heat kernel for the scalar field as a sum over images and the non-divergent part of the effective action. We restrict to the case of non-spinning BTZ geometry, or equivalently, thermal AdS with vanishing angular potential. Consider the indirect geodesic contributions for the $+$branch; as noticed by Perry and Williams~(\cite{PW03}, see also~\cite{BGW07}), this contribution can be written in terms of the Selberg zeta function
\be
W_{nondiv}=\mbox{ln\,Z}_{\Gamma}(\lambda).
\ee
The divergent part, when combined with the $-$branch, produces a finite factor (Plancherel measure) times the volume of the fundamental domain which vanishes in the present case~(\ref{vol}). Therefore, we get for the relative determinant  (remember that the effective action is just $W=\frac{1}{2} \mbox{ln\,det}_+(\Delta_X-\lambda(2-\lambda))$ )
\be
\frac{\mbox{det}_-(\Delta_X-\lambda(2-\lambda))}{\mbox{det}_+(\Delta_X-\lambda(2-\lambda))}
=\left[\mbox{Z}_{\Gamma}(2-\lambda)/\mbox{Z}_{\Gamma}(\lambda)\right]^2~,
\ee
in perfect agreement with the boundary result~(\ref{scat-n}).

\subsection{Heat kernel and Weyl's fractional calculus}

We proceed now to adapt the previous result from three to $n+1$ dimensions.
The plan is to generalize the computation in~\cite{MS96}, by inserting the heat kernel for the $(n+1)$-dimensional hyperbolic space.

The heat kernel $K=e^{-t\,\Delta_H}$ of the positive Laplacian in hyperbolic space satisfies a recurrence relation (see e.g.~\cite{GN98}). It is useful to treat it as a function of $x=\mbox{ch}\,\sigma$, where $\sigma$ is the geodesic distance, so that $x$ is essentially the chordal distance on the embedded hyperboloid~(\ref{chordal}):
\be
K_{n+2}(x,t)=-\frac{e^{-nt}}{2\pi}\frac{\partial}{\partial x}\,K_n(x,t)~,
\ee
The iteration begins with the one- and two-dimensional cases:
\bea
K_1&=\frac{1}{(4\pi t)^{1/2}}\,e^{-\sigma^2/4t}~,\\
K_2&=\frac{\sqrt{2}\,e^{-t/4}}{(4\pi t)^{3/2}}\,\int^{\infty}_{\sigma}ds\, \frac{s\,e^{-s^2/4t}}{(\mbox{ch}s-\mbox{ch}\sigma)^{1/2}}~.
\eea

One has to split into $n$ even and odd:
\bea
n=\mbox{even}\qquad K_{n+1}&=\frac{e^{-tn^2/4}}{(2\pi)^{n/2}}\,\left(-\frac{\partial}{\partial x}\right)^{n/2}\,K_1~,\\
n=\mbox{odd}\qquad K_{n+1}&=\frac{e^{-tn^2/4}}{(2\pi)^{[n/2]}}\,\left(-\frac{\partial}{\partial x}\right)^{[n/2]}\,K_2~,
\eea
where $[a]$ denotes the largest integer smaller than or equal to $a$.

Now, we notice that $K_2$ can be written as
\be
K_2=\frac{e^{-t/4}}{(2\pi)^{1/2}}\,\frac{1}{\pi^{1/2}}\int^{\infty}_x \frac{dy}{(y-x)^{1/2}}\,\left(-\frac{\partial}{\partial y}\right)\,K_1~,
\ee
and one can recognize Weyl's fractional derivative of order $1/2$ with respect to $x=\mbox{ch}\,\sigma$, i.e. $_{_x}W^{\frac{1}{2}}_{_\infty}$
in the notation of Miller and Ross ~\cite{MR93}.
Therefore, the recurrences can be solved in terms of Weyl's fractional derivative\be
\label{K}
K_{n+1}=\frac{e^{-tn^2/4}}{(2\pi)^{n/2}}\cdot\,_{_x}W^{\frac{n}{2}}_{_\infty}\left[K_1\right]~.
\ee
This unifying expression is not explicitly reported in~\cite{GN98}; however, we later learned~\cite{AO03} that these formulas are constantly being rediscovered.

\subsection{Indirect geodesic contributions: the elegant way}

Let us now examine the geodesic distance between image points, in ``scattering coordinates'', at $t=t', \widehat{\Omega}=\widehat{\Omega}', u=u'+Nl$
\be
\mbox{ch}\sigma_N=\mbox{ch}^2t\;\mbox{ch}Nl-\mbox{sh}^2t=1+2\mbox{ch}^2t\;\mbox{sh}^2Nl/2.
\ee
It turns out that the volume integral of a function depending only on this geodesic distance between image points
\be
\int d \mbox{vol}_X\, \bullet\;=l\,\mbox{vol(S}^{n-1})\int^{\infty}_0dt\,\mbox{ch}t\,\mbox{sh}^{n-1}t\,\bullet~,
\ee
after the change of integration variable from $t$ to $x=\mbox{ch}\sigma_N$
\be
l\,\mbox{vol(S}^{n-1})\frac{1}{2(\mbox{sh}^2Nl/2)^{n/2}}\int^{\infty}_{\mbox{ch}Nl}
\frac{dx}{(x-\mbox{ch}Nl)^{1-n/2}}\bullet~,
\ee
can be written as a Weyl's fractional integral
\be
\frac{l\,\pi^{n/2}}{(2\,\mbox{sh}^2Nl/2)^{n/2}}\cdot\, _{_{\mbox{ch}Nl}}W^{-\frac{n}{2}}_{_\infty}\bullet~.
\ee
Now, acting on the indirect contribution $K_{n+1}$~(\ref{K}), the convolution of the fractional integral with the fractional derivative~\footnote{We call this computation ``elegant'' because this is reminiscent of Abel's ``elegant'' solution to the tautochrone curve by using convolution of fractional integral  and derivative to solve the integral equation.} trivially results in the identity:
\be
\int d \mbox{vol}_X\, K_{n+1}\;=
\frac{l}{(4\,\mbox{sh}^2Nl/2)^{n/2}}\,e^{-tn^2/4}\,K_1|_{\sigma=Nl}~.
\ee
To get the Green's function it only remains to use the following Laplace transform to compute the proper-time integral
\be
\int^{\infty}_0 dt\,t^{-1/2}\,e^{-a/4t-pt}=(\pi/p)^{1/2}\,e^{-(ap)^{1/2}}~.
\ee
After the proper-time integral, we collect all indirect contributions to write
\be
(2\lambda-n)\mbox{tr}{(G^+_X-G^+_H)}
=2l\sum^{\infty}_{N=1}\frac{e^{-\lambda Nl}}{(1-e^{-Nl})^n}~.
\ee
A further integration in $\lambda$  and subtraction of the value at $n-\lambda$, finally brings us to the answer for the relative functional determinant
\be
\frac{\mbox{det}_+(\Delta_X-\lambda(n-\lambda))}{\mbox{det}_-(\Delta_X-\lambda(n-\lambda))}
=2\sum^{\infty}_{N=1}\,\frac{1}{N}\,\frac{e^{-(n-\lambda)Nl}-e^{-\lambda Nl}}{(1-e^{-lN})^n}~,
\ee
which in terms of the corresponding Selberg zeta function can be conveniently written, in total agreement with the boundary computation~(\ref{scat-n}), as

\be
\frac{\mbox{det}_-(\Delta_X-\lambda(n-\lambda))}{\mbox{det}_+(\Delta_X-\lambda(n-\lambda))}
=\left[\mbox{Z}_{\Gamma}(n-\lambda)/\mbox{Z}_{\Gamma}(\lambda)\right]^2~.
\ee

\section{Conclusion}

In this work we have explored the proposed holographic formula, as part of the AdS/CFT dictionary relating spacetime concepts in the bulk and field theory concepts on the boundary. For thermal AdS and for the non-spinning BTZ instanton, we have obtained explicit results encoded in the appropriate generalization of Selberg zeta function. For $n$ odd, our results agree with those of Guillarmou~\cite{Gui05} which were derived by other means and require certain definitions of the involved traces.
Let us mention that, in fact, the bulk computation follows from a theorem due to Patterson~\cite{Pat89} valid for the more general case of a hyperbolic (loxodromic) element, i.e. dilation and a rotation in Poincare coordinates~(\ref{loxo}). This general case includes the spinning BTZ instanton and calls for a corresponding explicit computation on the boundary.
However, the case of BTZ with a conical singularity (due to identification by an elliptic element of the isometry group, addressed in~\cite{MS96}) is not covered by the theorem so that it seems worthwhile to explore a further extension. Remarkably, there is already a Selberg zeta function attached to this conical defect geometry~\cite{Wil05}.

\ack I would like to gratefully acknowledge Rodrigo Aros for reading a preliminary version of the manuscript and for his valuable comments, and to Andreas Juhl for useful e-mail correspondence. I would also like to thank the staff of the Physics Department at the University Andr\'es Bello for the cordial hospitality. This work was funded through Fondecyt-Chile(Grant 3090012).

\section*{Appendix: Hyperbolic space}
\appendix
Hyperbolic space $H^{n+1}$ can be conveniently described as an immersed hyperboloid
\be
X^2_0+\mathbf{X}^2-X^2_{n+1}=-1
\ee
in $(n+2)$-dimensional Minkowski space,
\be
ds^2=dX^2_0+d\mathbf{X}^2-dX^2_{n+1}~.
\ee
The isometries are clearly those corresponding to the Lorentz group, with generators
\be
J_{AB}=X_A\frac{\partial}{\partial X^B}-X_B\frac{\partial}{\partial X^A}
\ee
with $A,B=0,1,...,n,n+1$.
The chordal distance $\Xi$ is related to the geodesic distance $\sigma$ in a way reminiscent of the spherical case
\be
\label{chordal}
\Xi^2=\parallel X-X'\parallel^2=2(\mbox{ch}\sigma-1)~.
\ee

Poincare half-space model is obtained via the following parametrization
\bea
\mathbf{X}&=\mathbf{x}/z ~,\\
X^+&=X^n+X^{n+1}=1/z~,\\
X^-&=X^n-X^{n+1}=-\frac{\mathbf{x}^2+z^2}{z}~.
\eea
The geodesic distance between points $(z,\mathbf{x})$ and $(z',\mathbf{x}')$ is then given by
\be
\label{xi}
\sigma(X,X')=\mbox{ln}(\frac{1+\sqrt{1-\xi^2}}{\xi})~,
\ee
where
\be
\xi=2zz'/[z^2+z'^2+(\mathbf{x}-\mathbf{x}')^2]~.
\ee

Among the isometries, the hyperbolic (or loxodromic) elements, which are relevant for the results by Patterson~\cite{Pat89}, are given by a dilation($\gamma$) combined with a rotation($\mathbb{A}$) in Poincare coordinates and parameterized by the pair $\{\gamma,\mathbb{A}\}$

\be
\label{loxo}
(z,\mathbf{x})\mapsto \gamma(z,\mathbb{A}\cdot \mathbf{x})~.
\ee
In this paper, we have merely considered the simplest case where $\mathbb{A}=\mathbb{I}$, i.e. no rotation in the transverse coordinates.
\setcounter{section}{1}

\end{document}